\documentclass{pramana}

\usepackage{graphicx,amsmath,bm}
\usepackage{soul}

\newcommand{\be}{\begin{equation}} \newcommand{\ee}{\end{equation}}
\newcommand{\bea}{\begin{eqnarray}} \newcommand{\eea}{\end{eqnarray}}
\newcommand{\bse}{\begin{subequations}} \newcommand{\ese}{\end{subequations}}
\newcommand{\n}{\nonumber}

\begin{document}

\title{A family of solutions to the Einstein-Maxwell system of equations describing relativistic charged fluid spheres}


\author{K. Komathiraj\textsuperscript{1}\and Ranjan Sharma\textsuperscript{2}}
\affilOne{\textsuperscript{1} Department of Mathematical Sciences, Faculty of Applied Sciences, South Eastern University, Sri Lanka.\\}
\affilTwo{\textsuperscript{2} Department of Physics, P. D. Women's College, Jalpaiguri 735101, India.}


\twocolumn[{

\maketitle

\corres{rsharma@associates.iucaa.in}


\begin{abstract}
In this paper, we present a formalism to generate a family of interior solutions to the Einstein-Maxwell system of equations for a spherically symmetric relativistic charged fluid sphere matched to the exterior Reissner-Nordstr\"om spacetime. By reducing the Einstein-Maxwell system to a recurrence relation with variable rational coefficients, we show that it is possible to obtain closed-form solutions for a specific range of the model parameters. A large class of solutions obtained previously are shown to be contained in our general class of solutions. We also analyze the physical viability of our new class of solutions.
\end{abstract}



}]



\section{Introduction}
Exact solutions to Einstein-Maxwell (EM) system of field equations play an important role in the studies of self-gravitating spherically symmmetric charged fluid distributions. Ever since the discovery of the Reissner-Nordstr\"om solution, many investigators have contributed to the study of EM system which includes the pioneering works of Papapetrou \cite{Pap}, Majumdar \cite{Maj}, Bonner \cite{Bon1, Bon2}, Stettner \cite{Ste}, Bekenstein \cite{Bek} and Cooperstock and Cruz \cite{Coo}. A detailed review of exact solutions to Einstein-Maxwell systems and their physical acceptability can be found in the compilation work of Ivanov \cite{Iv}. A large class of interior solutions, corresponding to the exterior  Reissner-Nordstr\"om  spacetime, have been developed to model a wide variety of stellar distributions such as neutron stars, strange stars and stellar objects composed of quark-diquark mixtures \cite{KoMa1, KoMa2,PaKo,TiSi,GuKu,ShMuMh,ShKaMu,ShMu1,ShMu2}. Stellar models have also been developed for charged core-envelope type configurations \cite{ThRaVi,TiTh,PaTi}. Mak and Harko \cite{MaHa} and Komathiraj and Maharaj \cite{KoMa3} have obtained solutions for charged strange quark stars admitting a linear equation of state (EOS). Thirukkanesh and Maharaj \cite{ThMa1} have analyzed the role of anisotropy on the physical behaviour of a given charged distribution admitting a linear EOS. Varela \textit{et al} \cite{Var} have analyzed features of a charged anisotropic fluid distribution admitting linear as well as non-linear EOS. Takisa and Maharaj \cite{TaMa} have obtained a new class of solutions for a charged quark matter distribution. Feroze and Siddiqui \cite{FeSi} and Maharaj and Takisa \cite{MaTa} have independently developed charged stellar models by assuming a quadratic EOS. Thirukkanesh and Ragel \cite{ThRa1, ThRa2} have obtained new solutions for charged fluid spheres by specifying the polytropic index leading to masses and energy densities which have been shown to be consistent with observational data. Maharaj \textit{et al} \cite{MaMa} have presented a new family of exact solutions to the Einstein-Maxwell system for an anisotropic charged matter on the Finch and Skea \cite{FiSk} background spacetime. A class of charged anisotropic stellar solutions has been developed and studied by Murad and Fatema \cite{MuFa}. Hansraj {\em et al} \cite{Hans_JMP} have analyzed all static-charged dust sphere models in general relativity. Recently, Sunzu and Danford \cite{Sunzu} have generated two new class of exact solutions to the Einstein-Maxwell system of field equations describing an anisotropic and charged stellar body which accommodates a quark matter like linear EOS.

The main objective of the present work is to contribute to this rich family of solutions by generating new solutions which can be used as viable models of realistic astrophysical objects. While generating the solutions, one needs to ensure that the gravitational, electromagnetic and matter variables remain finite, continuous, well behaved and the speed of sound remains less than the speed of light within the distribution. For a charged fluid sphere, the interior solution must be matched to the exterior Riessner-Nordstr{\"o}m metric across the boundary. We present here a different family of solutions to the coupled Einstein-Maxwell system where all the above requirements are fulfilled. This has been done by choosing a rational form for one of the gravitational potentials and also the fall-off behaviour of the charged fluid distribution. This particular approach is similar to the method adopted earlier by Maharaj and Leach \cite{MaLe} which was, in fact, a generalization of the superdense stellar model developed by Tikekar \cite{Tik}. In our approach, the solutions are generated by reducing the condition of pressure isotropy to a recurrence relation with real and rational coefficients so that the system can be solved by mathematical induction. We have performed a systematic analysis of the new family of solutions to examine their physical viability.  

The paper has been organized as follows: In Section $2$, we have presented the EM field equations for a static spherically symmetric charged fluid distribution. The nonlinear system was then transformed into a more tractable set of equations. By assuming a particular form for one of the metric potentials and also by specifying the electric field intensity, we have obtained the condition of pressure isotropy in terms of the undetermined gravitational potential in Section $3$. We have assumed a series solution for the resultant 
equation which yielded a recurrence relation. We have managed to solve the system from the first principles in Section $4$. In Section $5$,
we have presented polynomials and product of polynomials with algebraic functions as the first solution. The general solution containing the integral form was eventually integrated to yield elementary functions by placing specific restrictions on the model parameters. We have demonstrated that it is possible to regain many solutions found earlier by adopting this technique. Finally, we have provided two different class of exact solutions to the  EM system in simple closed forms. In Section $6$, we have discussed features of the class of solutions and showed that the solutions might be used to model realistic compact stellar systems. The results have been summarized in Section $7$.

\section{Einstein-Maxwell system}
For a static spherically symmetric relativistic charged fluid distribution, we assume the line element in coordinates $(t,r,\theta,\phi)$ as \be \label{eq:1}
ds^{2}=-e^{2\nu(r)}dt^{2}+e^{2\lambda(r)}dr^{2}+r^{2}(d\theta^{2}+\sin^{2}\theta
d\phi^{2}),\ee where $\nu(r)$ and $\lambda(r)$ are arbitrary functions of the radial coordinate $r$.  The Einstein-Maxwell
field equations for the line element (\ref{eq:1}) are then obtained (in system of units having 8$\pi G$ = $c$ = 1) as
\bse\label{eq:2}\bea
\frac{1}{r^{2}}(1-e^{-2\lambda})+\frac{2\lambda^\prime}{r}e^{-2\lambda}&=&\rho+\frac{1}{2}E^{2},\\
\frac{-1}{r^{2}}(1-e^{-2\lambda})+\frac{2\nu^\prime}{r}e^{-2\lambda}&=&p-\frac{1}{2}E^{2},\\
e^{-2\lambda}\left(\nu^{\prime\prime}+{\nu^\prime}^2+\frac{\nu^\prime}{r}-\nu^\prime\lambda^\prime-\frac{\lambda^\prime}{r}\right)&=&p+\frac{1}{2}E^{2},\\
\frac{1}{r^{2}}e^{-\lambda}(r^{2}E)^\prime&=&\sigma,\eea\ese 
where a prime ( $^{'}$) denotes differentiation with respect to $r$. The energy density $\rho$ and the pressure $p$ are measured relative
to the comoving fluid 4-velocity $u^{a}=e^{-\nu}\delta^{a}_{0}$. The electric field intensity $E$ and  the
proper charge density $\sigma$ appear into the system through the energy-momentum tensor corresponding to the electro-magnetic field
and the Maxwell equations.

A different but equivalent form of the field equations can be generated  if we introduce a new independent variable $x$ and
introduce new functions $y$ and $Z:$ 
\be \label{eq:3}
x=Cr^{2},~~~A^{2}y^{2}(x)=e^{2\nu(r)},~~~Z(x)=e^{-2\lambda(r)},
\ee
proposed by Durgapal and Bannerji \cite{DuBa}, where $A$ and $C$ are constants. Under the transformation (\ref{eq:3}),
the system (\ref{eq:2}) becomes
 \bse\label{eq:4} \bea
\label{eq:4a}\frac{1-Z}{x}-2\dot{Z} = \frac{\rho}{C}+\frac{E^{2}}{2C},\\
\label{eq:4b}4Z\frac{\dot{y}}{y}+\frac{Z-1}{x} = \frac{p}{C}-\frac{E^{2}}{2C},\\
\label{eq:4c}4Zx^{2}\ddot{y}+2\dot{Z}x^{2}\dot{y}+\left(\dot{Z}x-Z+1-\frac{E^{2}x}{C}\right)y = 0,\\
\label{eq:4d}\frac{4Z}{x}(x\dot{E}+E)^{2} = \frac{\sigma^{2}}{C},\eea\ese
where dots (.) denote differentiation with respect to the variable $x$. The system (\ref{eq:4}) determines the
gravitational behaviour of  a charged perfect fluid. Consequently, we have a nonlinear system  of four independent equations in six unknowns variables namely, $\rho$, $p$, $E$, $\sigma$, $y$ and $Z$. The advantage of this system lies in the fact that a solution,  upon  suitable substitutions of $Z$ and $E$, can be  obtained by integrating
 the second order differential equation (\ref{eq:4c}) which  is  linear in $y$.

\section{Integration procedure}
We solve the Einstein-Maxwell system (\ref{eq:4}) by making explicit choices for the metric function $Z$ and the electric field intensity $E$. For the metric function $Z$ we write 
\be\label{eq:5}Z=\frac{(1+k x)}{(1+mx)},~~~k\neq m,\ee 
where $k$ and $m$ are real constants. Note that the choice (\ref{eq:5}) ensures that the metric function $e^{2\lambda}$ is regular and
continuous in the interior because of the freedom provided by parameters $k$ and $m$. It is important to note that the particular choice
of $Z$ is physically reasonable and contains some special cases of known relativistic star models. The $m=1$ case corresponds to the
Maharaj and Komathiraj \cite{MaKo} charged stellar model which is a generalization of the stellar models developed previously by Finch and
Skea \cite{FiSk} and  Hansraj and Maharaj \cite{HaMa}. A similar form of $Z$ has also been utilized in Ref. \cite{FeSi, MaTa,
TaMa2} for the construction of a charged anisotropic stellar model admitting a polytropic EOS.

Substitution of (\ref{eq:5}) in (\ref{eq:4c}) yields
\bea\label{eq:6}4(1+kx)(1+mx)\ddot{y}+2(k-m)\dot{y}\nonumber\\+\left[m(m-k)-\frac{E^{2}(1+mx)^{2}}{Cx}\right]y=0.\eea
It is convenient at this point to introduce the following transformation \be\label{eq:7} z=1+mx.\ee This transformation
enables us to rewrite the second order differential equation (\ref{eq:6}) in a simpler form
\be\label{eq:8}4z(kz+m-k)\frac{d^{2}\tilde{y}}{dz^{2}}+2(k-m)\frac{d\tilde{y}}{dz}+\left[(m-k)-\frac{E^{2}z^{2}}{C(z-1)}\right]\tilde{y}=0,\ee
in terms of the new dependent and independent variables $\tilde{y}=y(z)$ and $z$, respectively. To integrate (\ref{eq:8}),
it is necessary to specify the electric field intensity $E$. Even though a variety of choices for $E$ is possible,  only a few of
them are  physically reasonable and can  generate closed form solutions. We  reduce (\ref{eq:8}) to an integrable  form by
letting
\be\label{eq:9}\frac{E^{2}}{C} = \alpha\frac{(z-1)}{z^{2}}-\beta\frac{(z-1)}{z^{3}} = \alpha \frac{ mx}{(1+mx)^{2}}-\beta \frac{ mx}{(1+mx)^{3}},\ee
where $\alpha$ and $\beta$ are  constants. The form $E^{2}$ in (\ref{eq:9}) is physically acceptable as $E$ remains regular and continuous throughout the sphere. Note that $E=0$ at $r=0$.  Some special cases of (\ref{eq:9}) have earlier been studied by  Takisa and Maharaj \cite{TaMa2} and John and Maharaj \cite{JoMa} and can also be reduced to the uncharged  stellar model developed by Maharaj and Mkhwanazi \cite{MaMk}. Substituting (\ref{eq:9}) in equation (\ref{eq:8}), we obtain 
\be\label{eq:10}4z^{2}(kz+m-k)\frac{d^{2}\tilde{y}}{dz^{2}}+2z(k-m)\frac{d\tilde{y}}{dz}+[(m-k-\alpha)z+\beta]\tilde{y}=0,\ee
which is the master equation for the system of equations (\ref{eq:4}). For $\alpha=\beta=0$, the differential equation
(\ref{eq:10}) reduces to
\be\label{eq:11}4z(kz+m-k)\frac{d^{2}\tilde{y}}{dz^{2}}+2(k-m)\frac{d\tilde{y}}{dz}+(m-k)\tilde{y}=0,\ee
which is the limiting case and corresponds to an uncharged sphere.

\section{General series solution}
A closed form solution of  equation (\ref{eq:10}) is difficult to obtain. However, one can transform it to a  differential equation
which can be integrated by the method of Frobenius. This can be done in the following way.

We introduce a new function $u(z)$ such that \be
\label{eq:12}\tilde{y}(z)=z^{d}u(z),\ee 
where $d$ is a constant. A similar kind of transformation was utilized earlier by Komathiraj and Maharaj \cite{KoMa4} for generating  charged stellar models. With the help of (\ref{eq:12}), the differential equation (\ref{eq:10}) can be written as \bea \label{eq:13}
&&4z^{2}(kz+m-k)\frac{d^{2}u}{dz^{2}}\n\\
&-&+2z[4d(kz+m-k)+(k-m)]\frac{du}{dz}\n\\
&-&\{[k-m+\alpha-4kd(d-1)]z\n\\
&-& +2d(2d-3)(k-m)-\beta\}u=0.\eea 
A substantial simplification of the equation can be achieved if we set \be \label{eq:14} 2d(2d-3)(k-m)=\beta .\ee 
Equation (\ref{eq:13}) then reduces to  
\be \label{eq:15}
4z(z-c)\frac{d^{2}u}{dz^{2}}+2[4d(z-c)+c]\frac{du}{dz}-[c+\tilde{\alpha}-4d(d-1)]u=0,\ee
where we have set \be \label{eq:16} c=\frac{k-m}{k},~~~~~k
\tilde{\alpha}=\alpha . \ee 
Since the point $z=c$ is a regular singular point of (\ref{eq:15}), there exists two linearly independent solutions of the form of power series with centre $z=c$. Therefore, we can write  the solution of the differential equation (\ref{eq:15}) by the method of Frobenius as:
\be\label{eq:17}u=\sum_{i=0}^{\infty}c_{i}(z-c)^{i+b},~~~
c_{0}\neq0 ,\ee\\where $c_{i}$ are the coefficients of the series and $b$ is a constant.

For a legitimate  solution, we need to determine the coefficients $c_{i}$ as well as the parameter $b$. Substituting (\ref{eq:17})
in the differential equation (\ref{eq:15}), we obtain the indicial equation  \be 2cc_{0}b(2b-1)=0, \n\ee and the recurrence formula
\be\label{eq:18}
c_{i+1}=-\frac{\{4(i+b)(i+b-1+2d)-T\}}
{2c(i+b+1)(2i+2b+1)}c_{i},\ee 
with $i\geq 0$ and $T=c+\tilde{\alpha}-4d(d-1)$. Since $c_{0}\neq 0$, $c=k-m\neq 0$, we must have $b=0$ or $b=\frac{1}{2}$. The coefficients $c_{1},~c_{2}, c_{3},~.~.~.$ can all be written in
terms of the leading coefficient $c_{0}$ and we can generate the expression as
\be \label{eq:19}
c_{i+1}=\prod_{p=0}^{i}-\frac{4(p+b)(p+b-1+2d)-T}{2c(p+b+1)(2p+2b+1)}c_{0}.
\ee
It is also possible to establish the result  (\ref{eq:19}) rigorously by using the principle of mathematical induction.

We  generate two linearly independent solutions to (\ref{eq:15}) with the help of  (\ref{eq:17}) and (\ref{eq:19}). For the
parameter value $b=0$, we obtain the first solution 
\be
u_{1}=c_{0}\left[1+\sum_{i=0}^{\infty}\prod_{p=0}^{i}-\frac{\psi}{\eta}(z-c)^{i+1}\right],
\n \ee 
where $\psi=4p(p-1+2d)-[c+\tilde{\alpha}-4d(d-1)],$
and $\eta=2c(p+1)(2p+1).$
For the parameter value $b=\frac{1}{2}$, we obtain the second solution 
\be
u_{2}=c_{0}(z-c)^{\frac{1}{2}}\left[1+\sum_{i=0}^{\infty}\prod_{p=0}^{i}-\frac{\alpha}{\beta}(z-c)^{i+1}\right],\n\ee
where $\alpha=(2p+1)(2p-1+4d)-[c+\tilde{\alpha}-4d(d-1)],$ and  $\beta=c(2p+3)(2p+2).$
We, therefore, have a general solution  to (\ref{eq:15}) as the functions $u_{1}$ and $u_{2}$ are linearly independent. In terms
of the original variable $x=Cr^{2}$, the functions are obtained as 
\be \label{eq:20} y_{1}=
c_{0}(1+mx)^{d}\left[1+\sum_{i=0}^{\infty}\prod_{p=0}^{i}-\frac{\psi_1}{\eta_1}\left[\frac{m(1+kx)}{k}\right]^{i+1}\right],\ee
\bea  \label{eq:21}
y_{2}&=&c_{0}(1+mx)^{d}\left[\frac{m(1+kx)}{k}\right]^{\frac{1}{2}}\n\\
&\times&\left[1+\sum_{i=0}^{\infty}\prod_{p=0}^{i}-\frac{\alpha_1}{\beta_1}\left[\frac{m(1+kx)}{k}\right]^{i+1}\right],\eea
where
$$\psi_1=4kp(p-1+2d)-[k-m+\alpha-4kd(d-1)],$$ $$\eta_1=2(k-m)(p+1)(2p+1),$$ $$\alpha_1=k(2p+1)(2p-1+4d)-[k-m+\alpha-4kd(d-1)],$$  
$$\beta_1=(k-m)(2p+3)(2p+2).$$
Thus the general solution to the differential equation
(\ref{eq:6}), for the choice of the electric field (\ref{eq:9}), is given by \be \label{eq:22}y(x)=A_{1}y_{1}(x)+A_{2}y_{2}(x).\ee
where $A_{1}$ and $A_{2}$ are arbitrary constants. Using,  (\ref{eq:4}) and (\ref{eq:22}), we  write the exact solution to the Einstein-Maxwell system in the
form
\bse\label{eq:23}\bea\label{eq:23a}e^{2\lambda}&=&\frac{1+mx}{1+kx},\\
\label{eq:23b}e^{2\nu}&=&A^{2}y^{2},\\
\frac{\rho}{C}&=&\frac{(m-k)(3+mx)}{(1+mx)^{2}}-\alpha\frac{mx}{2(1+mx)^{2}}\n\\
&&+\beta\frac{mx}{2(1+mx)^{3}},\\
\frac{p}{C}&=&4\frac{(1+kx)}{(1+mx)}\frac{\dot{y}}{y}+\frac{k-m}{(1+mx)}\n\\&&+\alpha\frac{mx}{2(1+mx)^{2}}-\beta\frac{mx}{2(1+mx)^{3}},\\
\frac{E^{2}}{C}&=&\alpha\frac{mx}{(1+mx)^{2}}-\beta\frac{mx}{(1+mx)^{3}}.\eea\ese
As the choice of the metric function (\ref{eq:5}) together with the electric field intensity (\ref{eq:9}) have not been considered
earlier, to the best of our knowledge, the class of  solutions (\ref{eq:23}) have not been reported previously. One interesting
feature of the new family of solutions is that by setting $\alpha=0$ and $\beta=0~(d=0~\textrm{or}~d=\frac{3}{2})$, it is possible to switch
off the effect of charge onto the system. Secondly, the solution (\ref{eq:22}) has been expressed in
terms of a series of real arguments and not complex arguments which one might encounter when mathematical software packages are used.

\section{Terminating series}
It is interesting to observe that the series in (\ref{eq:20}) and (\ref{eq:21}) terminates for specific  values of the parameters
$k,~m,~\alpha$ and $d$. It is, therefore,   possible to generate solutions in terms of elementary functions by imposing  specific
restrictions on $k,~m,~\alpha$ and $d$. The solutions may be found in terms of polynomials and algebraic functions.   We use
recurrence relation (\ref{eq:18}), rather than the series (\ref{eq:20}) and (\ref{eq:21}), to find the elementary solutions.

\subsection{Elementary solutions}
If we fix $b=0$ in (\ref{eq:18}), and set $c+\tilde{\alpha}-4d(d-1)=4n(n-1+2d)$, for integer values of $n$,
we obtain
\be\label{eq:24}c_{i+1}=4\frac{(n-i)(n+i-1+2d)}{c(2i+1)(2i+2)}c_{i},~i\geq0,\ee
where $n$ is a fixed integer. Obviously, $c_{n+1}=0$. Consequently, the remaining coefficients
$c_{n+2},~c_{n+3},~c_{n+4},~.~.~.$ vanish. Equation (\ref{eq:24}) may be solved to yield
\be\label{eq:25}c_{i}=\left(\frac{4}{c}\right)^{i}\frac{n!(n+i-2+2d)!}{(2i)!(n-i)!(n-2+2d)!}c_{0},~0\leq
i\leq n.\ee 
Using (\ref{eq:17}) (when $b=0$) and (\ref{eq:25}), we obtain
\be\label{eq:26}u(z)=c_{0}\sum_{i=0}^{n}\left(\frac{4}{c}\right)^{i}\frac{n!(n+i-2+2d)!}{(2i)!(n-i)!(n-2+2d)!}(z-c)^{i},\ee
where $c+\tilde{\alpha}-4d(d-1)=4n(n-1+2d)$.

On substituting $b=\frac{1}{2}$ in (\ref{eq:18}) and by setting $c+\tilde{\alpha}-4d(d-1)=(2n+1)(2n-1+4d)$, we
obtain\be\label{eq:27}c_{i+1}=\frac{4(n-i)(n+i+2d)}{c(2i+3)(2i+2)}c_{i},~i\geq0,\ee
where $n$ is a fixed integer. Obviously, $c_{n+1}=0$ and the subsequent coefficients
$c_{n+2},$ $c_{n+3},$ $c_{n+4},~.~.$ vanish.  Equation (\ref{eq:27}) yields
\be\label{eq:28}c_{i}=\left(\frac{4}{c}\right)^{i}\frac{(n)!(n+i-1+2d)!}{(2i+1)!(n-i)!(n-1+2d)!}c_{0},~0\leq
i\leq n.\ee 
Using (\ref{eq:17}) (when $b=\frac{1}{2}$) and (\ref{eq:28}), we obtain
\be\label{eq:29}u(z)=c_{0}(z-c)^{\frac{1}{2}}\sum_{i=0}^{n}\left(\frac{4}{c}\right)^{i}\frac{(n)!(n_1+i)!}{(2i+1)!(n-i)!n_1!}(z-c)^{i},\ee
where $n_1=(n-1+2d)$ and $c+\tilde{\alpha}-4d(d-1)=(2n+1)(2n-1+4d)$.
The polynomial (\ref{eq:26}) and the product of polynomial and algebraic function (\ref{eq:29}) generate a particular solution of
the differential equation (\ref{eq:15}) for appropriate values of the parameters $c,~\tilde{\alpha}$ and $d$.

\subsection{General solutions} 
It is possible to obtain  solutions to (\ref{eq:15}) by restricting the values of $c,~\tilde{\alpha}$ and $d$ so that only elementary functions survive. The elementary functions are expressible as polynomials and product of polynomials with algebraic functions. Using
(\ref{eq:26}), we  express the first category of general solutions to the differential equation  (\ref{eq:15}) in the form
\bea\label{eq:30}u(z)=\sum_{i=0}^{n}\left(\frac{4}{c}\right)^{i}\frac{n!(n+i-2+2d)!}{(2i)!(n-i)!(n-2+2d)!}(z-c)^{i}\n\\
\times \left[B_{1}+B_{2}\int
\frac{z^{\frac{1}{2}-2d}(z-c)^{-\frac{1}{2}}}{\{\sum_{i=0}^{n}\left(\frac{4}{c}\right)^{i}\frac{n!(n_2+i)!}{(2i)!(n-i)!n_2!}(z-c)^{i}\}^{2}}dz\right
],\eea where $n_2=(n-2+2d)$ and $c+\tilde{\alpha}-4d(d-1)=4n(n-1+2d)$. 
Using(\ref{eq:29}), the second category of general solutions to (\ref{eq:15}) is obtained as
\bea\label{eq:31}u(z)=(z-c)^{\frac{1}{2}}\sum_{i=0}^{n}\left(\frac{4}{c}\right)^{i}\frac{n!(n_1+i)!}{(2i+1)!(n-i)!n_1!}(z-c)^{i}\n\\
\times \left[B_{1}+B_{2}\int
\frac{z^{\frac{1}{2}-2d}(z-c)^{-\frac{3}{2}}}{\{\sum_{i=0}^{n}\left(\frac{4}{c}\right)^{i}\frac{n!(n_1+i)!}{(2i+1)!(n-i)!n_1!}(z-c)^{i}\}^{2}}dz\right],\eea
where $c+\tilde{\alpha}-4d(d-1)=(2n+1)(2n-1+4d)$. In (\ref{eq:30}) - (\ref{eq:31}), $B_{1}$ and $B_{2}$  are integration
constants. In terms of the original variable $x=Cr^{2}$, it is possible to write  (\ref{eq:30}) in the form
\bea\label{eq:32}y(x)= (1+mx)^{d}\sum_{i=0}^{n}\left(\frac{4k}{k-m}\right)^{i}\frac{n!(n_2+i)!}{(2i)!(n-i)!n_2!}t^{i}\n\\
\times \left[C_{1}+C_{2}\int
\frac{(1+mx)^{\frac{1}{2}-2d}[(1+kx)]^{-\frac{1}{2}}}{\{\sum_{i=0}^{n}\left(\frac{4k}{k-m}\right)^{i}\frac{n!(n_2+i)!}{(2i)!(n-i)!n_2!}[\frac{m(1+kx)}{k}]^{i}\}^{2}}dx\right],
\eea where  $k-m+\alpha-4kd(d-1)=4kn(n-1+2d)$.
Equation (\ref{eq:31}) in terms of $x=Cr^{2}$, takes the form
\bea\label{eq:33}y(x)=C_3\sum_{i=0}^{n}\left(\frac{4k}{k-m}\right)^{i}\frac{n!(n_1+i)!}{(2i+1)!(n-i)!n_1!}t^{i}\n\\
\times\left[C_{1}+C_{2}\int
\frac{(1+mx)^{\frac{1}{2}-2d}[(1+kx)]^{-\frac{3}{2}}}{\{\sum_{i=0}^{n}\left(\frac{4k}{k-m}\right)^{i}\frac{n!(n_1+i)!}{(2i+1)!(n-i)!n_1!}t^{i}\}^{2}}dx\right],\eea
where $k-m+\alpha-4kd(d-1)=k(2n+1)(2n-1+4d),$ $C_3=(1+mx)^{d}(1+kx)^{\frac{1}{2}}$ and $t=m(1+kx)/k$.

We, thus have generated two class of solutions  (\ref{eq:32}) and (\ref{eq:33}) to the differential equation (\ref{eq:6}) for the
assumed electric field (\ref{eq:9}) making use of the infinite series solution (\ref{eq:22}). It should be stressed that the class of
solutions can be used to study stellar properties in the presence as well as absence of charge. By setting $\alpha=0$ and $\beta=0~(d=0 ~\textrm{or}~\frac{3}{2})$ in (\ref{eq:32}) and (\ref{eq:33}), one obtains solutions for an uncharged sphere.

We are now in a position to integrate equations (\ref{eq:32}) and (\ref{eq:33}) for specific values of the
parameters $k,~m,~\alpha, d$ and $n$.

For $n=0$, equation (\ref{eq:32}) becomes
\be \label{eq:34}
y(x)=(1+mx)^{d}\left[C_{1}+C_{2}\int\frac{1}{(1+mx)^{2d-\frac{1}{2}}(1+kx)^{\frac{1}{2}}}dx\right],\ee
where $k-m+\alpha-4kd(d-1)=0$. Also for $n=0$, equation (\ref{eq:33})
takes the form  \be \label{eq:35}
y(x)=C_3\left[C_{1}+C_{2}\int\frac{1}{(1+mx)^{2d-\frac{1}{2}}(1+kx)^{\frac{3}{2}}}dx\right],\ee
where $2k-m+\alpha-4kd^{2}=0$.

By setting $k=-\frac{1}{2},~ m=1,~
\alpha=0 $ and $d=\frac{3}{2}~(\textrm{or}~ \beta =0)$ in
(\ref{eq:34}), we obtain \be
y(x)=a_{1}(1+x)^{\frac{3}{2}}+a_{2}(2-x)^{\frac{1}{2}}(5+2x),\n\ee
where we have assumed $a_{1}=C_{1}$ and
$a_{2}=-\frac{2\sqrt{2}}{27}C_{2}$. Thus, we have regained the Durgapal and Bannerji solution \cite{DuBa}.

If we set $k=\frac{1}{2},~ m=1,~ \alpha=0 $ and $d=0~(\textrm{or}~
\beta =0)$ in (\ref{eq:35}),  we obtain \be
y(x)=(2+x)^{\frac{1}{2}}(a_{1}+a_{2}\ln[(1+x)^{\frac{1}{2}}+(2+x)^{\frac{1}{2}}])-a_{2}(1+x)^{\frac{1}{2}},\n\ee
where we have assumed $a_{1}=\frac{C_{1}}{\sqrt{2}}$ and $a_{2}=4C_{2}$. This class of solutions  was found earlier by
Maharaj and Mikhwanazi \cite{MaMk}.

Further, by setting $~k=\frac{1}{3},~m=1,~\alpha=\frac{1}{3}$ and $d=0~(\textrm{or}~ \beta =0)$ in   (\ref{eq:35}),  we obtain
\be
y(x)=(3+x)^{\frac{1}{2}}(a_{1}+a_{2}\ln[(1+x)^{\frac{1}{2}}+(3+x)^{\frac{1}{2}}])-a_{2}(1+x)^{\frac{1}{2}},\n\ee
where we have assumed  $a_{1}=\frac{C_{1}}{\sqrt{3}}$ and $a_{2}=6C_{2}$. This particular solution corresponds to
the charged stellar model of John and Maharaj \cite{JoMa}. Note that a minor error appearing in the John and Maharaj paper \cite{JoMa}
has been addressed in this work.

\subsection{New family of solutions} 
We now aim to generate new closed form solutions for $y$ which can subsequently be used to model realistic stars. To achieve our objective we set $k=-\frac{1}{2},~ m=2$ and $\alpha=1$. Then, making use of (\ref{eq:34}), it is possible to obtain  two categories of solutions for (i) $d=\frac{3}{2} (\beta=0)$ and (ii) $d=-\frac{1}{2}~(\beta=-10).$ \\
\textbf{Case I:}  $d=\frac{3}{2}~(\beta=0)$\\
In this case (\ref{eq:34}) becomes
\be \label{36}
y(x)=a_{1}(1+2x)^{\frac{3}{2}}+a_{2}(2-x)^{\frac{1}{2}}(7+4x),\ee
 where we have set $a_{1}=C_{1},~a_{2}=-\frac{2}{75}C_{2}.$

Subsequently, the general solution to the Einstein-Maxwell system (\ref{eq:23}) can be  expressed as \bse\label{37}\bea
e^{2\lambda}&=&\frac{2(1+2x)}{2-x},\\
e^{2\nu}&=&A^{2}\{a_{1}(1+2x)^{\frac{3}{2}}\nonumber\\
&&+a_{2}(2-x)^{\frac{1}{2}}(7+4x)\}^{2},\\
\label{37c}\frac{\rho}{C}&=&\frac{15+8x}{2(1+2x)^{2}},\\
\label{37d}\frac{p}{C}&=&\frac{g_1(x)}{g_2(x)},\n\\\\
\frac{E^{2}}{C}&=&\frac{2x}{(1+2x)^{2}}, \eea\ese
where $g_1(x)=a_{1}(2-x)^{\frac{1}{2}}(1+2x)^{\frac{1}{2}}(19+18x-40x^{2})+a_{2}(-34-111x-96x^{2}+80x^{3})$ and $g_2(x)=2(2-x)^{\frac{1}{2}}(1+2x)^{2}[a_{1}(1+2x)^{\frac{3}{2}}+a_{2}(2-x)^{\frac{1}{2}}(7+4x)].$

\textbf{Case II:}  $d=-\frac{1}{2}~(\beta=-10)$\\
In this case (\ref{eq:34}) becomes \bea \label{eq:38}
y(x)=a_{1}(1+2x)^{-\frac{1}{2}}+a_{2}\left[-2\sqrt{2-x}(17+4x)\right.\n\\
\left.-75\sqrt{2}(1+2x)^{-\frac{1}{2}}\arcsin\left(\sqrt{\frac{2(2-x)}{5}}\right)\right],\eea
where we have set $a_{1}=C_{1},~a_{2}=\frac{1}{8}C_{2}.$

The simple form of our class of solutions facilitates the analysis of matter and gravitational variables of realistic stellar objects as can be seen in the following sections.

\section{Physical analysis}
For a physically viable model, the class of solutions obtained by our approach should satisfy certain regularity and physical requirements \cite{DeLa}. In this section, we analyze the features of our solutions and examine whether the solutions can be used for the description of realistic stars.

Note that we should restrict our solutions only to those values of $k$ and $m$ for which the energy density $\rho$, pressure $p$
and the electric field intensity $E$ remain finite and positive. In addition, $k$ and $m$ should be so chosen  that the gravitational
potential $ e^{2\lambda}$ remains positive since the other metric function $e^{2\nu}$ is obviously positive. In (\ref{eq:23a})
and (\ref{eq:23b}), we note that $ e^{2\lambda}$ and $e^{2\nu}$ are continuous in the stellar interior. They are also regular at the centre for
all values of the parameters $k,~m,~\alpha$ and $d~(\textrm{or}~\beta)$. 

That pressure of a realistic star must vanish at a finite boundary $r=R$ implies that
\bea
4(1+kCR^{2})\left[\frac{\dot{y}}{y}\right]_{R}+k-m+\frac{\alpha
mCR^{2}}{2(1+mCR^{2})}\n\\
-\frac{\beta mCR^{2}}{2(1+mCR^{2})^{2}}=0,
\n\eea where $y$ is given by (\ref{eq:22}). The above equation puts a restriction on the constants  $A_{1}$ and $A_{2}$. 

The unique solution to the Einstein-Maxwell system for $r > R$ is given by the Riessner-Nordstr{\"o}m  line element \bea \label{eq:39}
ds^{2}&=&-\left(1-\frac{2M}{r}+\frac{Q^{2}}{r^{2}}\right)dt^{2}+\left(1-\frac{2M}{r}+\frac{Q^{2}}{r^{2}}\right)^{-1}dr^{2}\n\\
&&+r^{2}(d\theta^{2}+\sin^{2}\theta
d\phi^{2}),\eea where $M$ and $Q$ are the total mass and the charge of the star. Matching of  the line elements (\ref{eq:1}) and (\ref{eq:39}), across  the boundary $r=R$,  yields the relationships between the constants  $A_{1},~A_{2},~k,~m$ and $R$ as follows:
 \bea
1-\frac{2M}{R}+\frac{Q^{2}}{R^{2}}&=&A^{2}[A_{1}y_{1}(CR^{2})+A_{2}y_{2}(CR^{2})]^{2},\n\\\n\\
\left(1-\frac{2M}{R}+\frac{Q^{2}}{R^{2}}\right)^{-1}&=&\frac{1+mCR^{2}}{1+kCR^{2}}.\n\eea

For the particular solution, using (\ref{37c}), we obtain the central density $$\rho_{0}=\rho(r=0)=\frac{15}{2}C,$$ which implies that $C>0$. To
obtain bounds on other parameters, we evaluate the pressure at two different points. Using equation (\ref{37d}) at the centre of the star ($r=0$),  we obtain the central pressure \be
p_{0}=p(x=0)=\frac{C(19a_{1}-17\sqrt{2}a_{2})}{2(a_{1}+7\sqrt{2}a_{2})}.\n\ee
Obviously, we must have \be \label{40} \frac{(19a_{1}-17\sqrt{2}a_{2})}{(a_{1}+7\sqrt{2}a_{2})} >0,\ee as $\frac{C}{2}>0$.\\
At the boundary of the star ($r=R$), we impose the condition that the pressure vanishes, i.e.,~$p_{R}=p(x=CR^{2})=0$, which yields
\bea
\label{41}a_{1}\sqrt{(2-CR^{2})(1+2CR^{2})}(19+18CR^{2}-40C^{2}R^{4})\n\\+a_{2}(-34-111CR^{2}-96C^{2}R^{4}+80C^{3}R^{6})=0.
\eea Equation (\ref{41}) determines the radius $R$ of the star. 

From (\ref{37c}), we note that the density is always positive and
\be \label{42} \frac{d\rho}{dr}=-\frac{4C^{2}r(13+4Cr^{2})}{(1+2Cr^{2})^{3}} <0.\ee 
We must also have \be \label{43}
\frac{dp}{dr}=-\frac{C^{2}r[-8a_{1}^{2}~f(r)+a_{1}a_{2}~g(r)+8a_{2}^{2}~h(r)]}{\sqrt{(2-Cr^{2})}(1+2Cr^{2})^{\frac{7}{2}}n_1}<0,\ee
where $n_1=[a_{1}(1+2Cr^{2})^{\frac{3}{2}}+a_{2}\sqrt{2-Cr^{2}}(7+4Cr^{2})]^{2}.$
To fulfill the causality condition $0<\frac{dp}{d\rho}<1$, we must have \be \label{44}
0<\frac{-8a_{1}^{2}~f(r)+a_{1}a_{2}~g(r)+8a_{2}^{2}~h(r)}{H(r)}<1,\ee
throughout the interior of the star where \bea f(r)&=&\sqrt{2-Cr^{2}}(1+2Cr^{2})^{\frac{7}{2}}(-12+5Cr^{2}),\n\\
g(r)&=&(1+2Cr^{2})^{2}(513-28Cr^{2}-848C^{2}r^{4}+320C^{3}r^{6}),\n\\
h(r)&=&\sqrt{(2-Cr^{2})(1+2Cr^{2})}(-24-34Cr^{2}-153C^{2}r^{4}\n\\
&&-72C^{3}r^{6}+80C^{4}r^{8}),\n\\
H(r)&=&4\sqrt{(2-Cr^{2})(1+2Cr^{2})}(13+4Cr^{2})[a_{1}(1+2Cr^{2})^{\frac{3}{2}}\n\\
&&+a_{2}\sqrt{2-Cr^{2}}(7+4Cr^{2})]^{2}.
\eea
It is not difficult to note that, at the centre of the star $(x=0)$, the causality condition puts a constraint
\be\label{45}0<\frac{96\sqrt{2}a_{1}^{2}+513a_{1}a_{2}-192\sqrt{2}a_{2}^{2}}{52\sqrt{2}(a_{1}+7\sqrt{2}a_{2})^{2}}<1.\ee

Using the matching conditions, we also obtain
\bea\label{46}  1-\frac{2M}{R}+\frac{Q^{2}}{R^{2}}&=&A^{2}\left[a_{2}(2-CR^{2})^{\frac{1}{2}}(7+4CR^{2})\right.\n\\
&&\left.+ a_{1}(1+2CR^{2})^{\frac{3}{2}}\right]^{2},\\
\label{47}\left(1-\frac{2M}{R}+\frac{Q^{2}}{R^{2}}\right)^{-1}&=&\frac{2(1+2CR^{2})}{2-CR^{2}}.\eea
Equation (\ref{46}) determines the values of the constants in terms of the total mass $M$, radius $R$ and the charge $Q$. From
equation (\ref{47}), we obtain the total mass of the star as  \be
M=\frac{CR^{3}(5+14CR^{2})}{4(1+2CR^{2})^{2}}.\n\ee
Since all the parameters on the right hand side of this equation have positive values, the mass of the star is finite and positive.

\begin{figure}
\centering\includegraphics[height=.2\textheight]{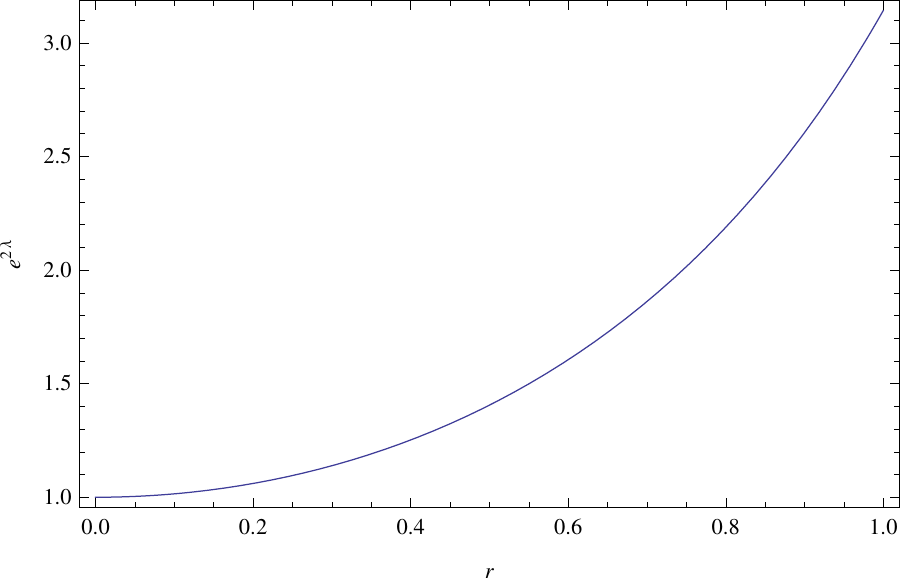}
\caption{Behaviour of gravitational potential $e^{\lambda}$.}\label{fig:1}
\end{figure}

\begin{figure}
\centering\includegraphics[height=.2\textheight]{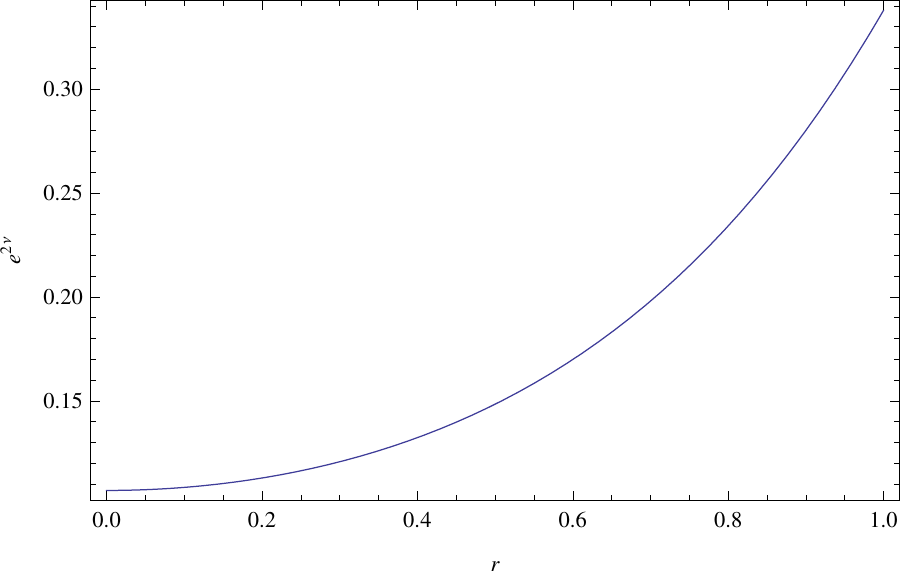}
\caption{Behaviour of gravitational potential $e^{\nu}$.}\label{fig:2}
\end{figure}

\begin{figure}
\centering\includegraphics[height=.2\textheight]{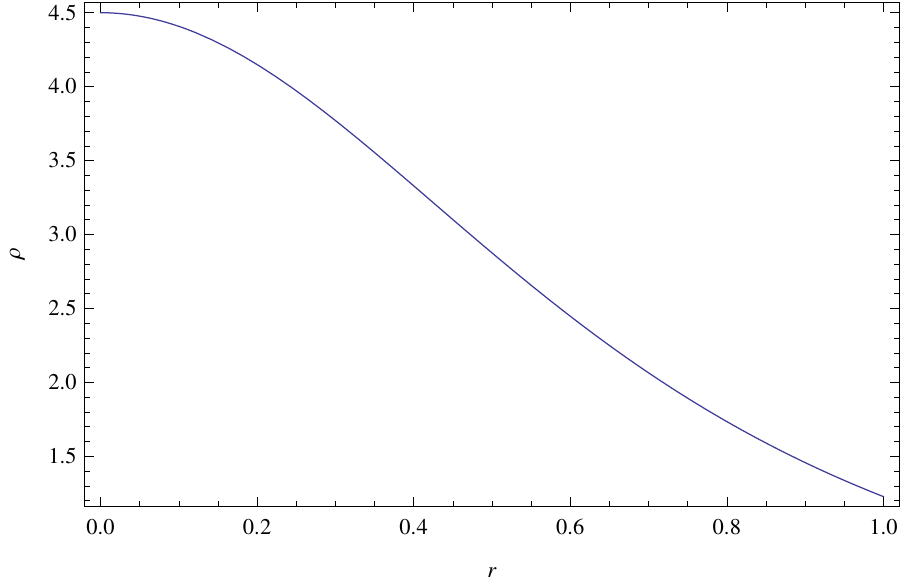}
\caption{Radial variation of density $\rho$.}\label{fig:3}
\end{figure}

\begin{figure}
\centering\includegraphics[height=.2\textheight]{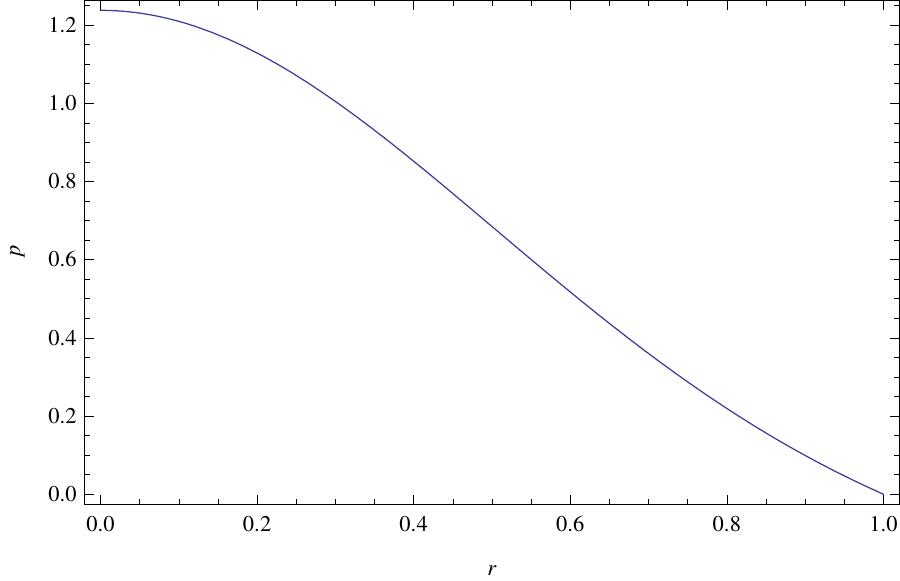}
\caption{Radial variation of pressure $p$.}\label{fig:4}
\end{figure}

\begin{figure}
\centering\includegraphics[height=.2\textheight]{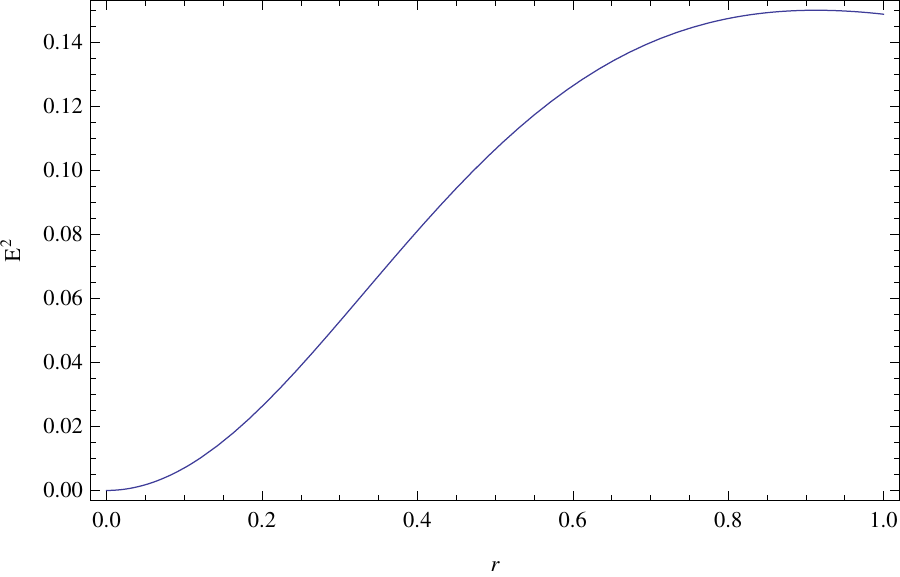}
\caption{Radial variation of electric field $E^2$.}\label{fig:5}
\end{figure}

\begin{figure}
\centering\includegraphics[height=.2\textheight]{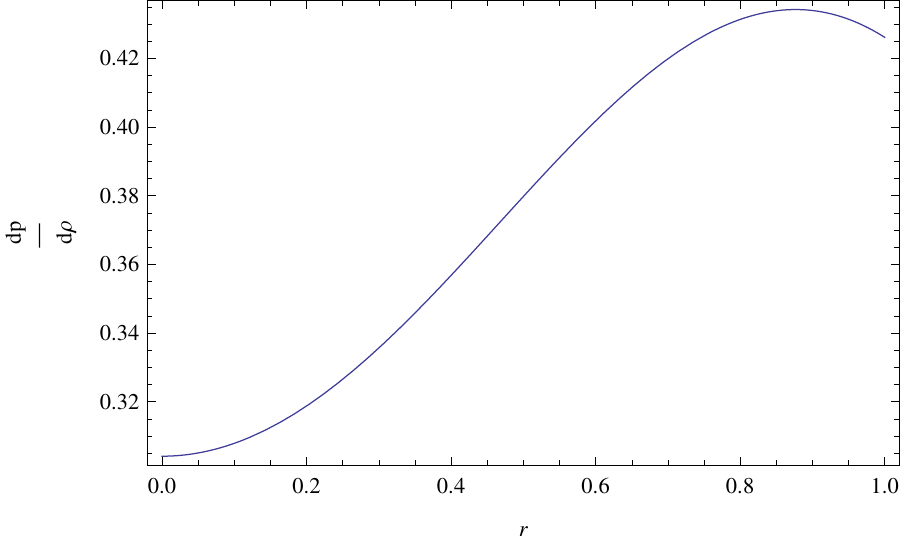}
\caption{Radial variation of square of sound speed $\frac{dp}{d\rho}$.}\label{fig:6}
\end{figure}

To analyze physical behaviour of a star, we have set $a_{1}=1,~R=1,~C=0.6$ which are consistent with the bounds (\ref{40})-(\ref{45}).  Using these values in equation (\ref{41}), we have obtained $a_{2}=0.229275$. In Fig.~\ref{fig:1} and \ref{fig:2}, we have plotted the gravitational potentials which have been shown to be well behaved. The behaviour of the energy density and the isotropic pressure have been shown in Fig.~\ref{fig:3} and \ref{fig:4}, respectively. We note that the energy density and pressure are positive and monotonically decreasing within the stellar interior and the pressure vanishes at the boundary. In Fig.~\ref{fig:5}, we have shown the fall-off
behaviour of the electric field intensity $E$. In Fig.~\ref{fig:6}, we have plotted $\frac{dp}{d \rho}$ on the
interval $0\leq r \leq 1$. We note that $\frac{dp}{d \rho}$ is always positive and less than unity, i.e., causality condition is
not violated. It,  therefore, can be concluded  that there exists particular set of model parameters for which the solution
(\ref{37}) satisfies all the requirements of a realistic star.

Another interesting feature of our model is that it allows us to generate a barotropic relationship between the energy density and pressure. A closed form thermodynamics relationship between density and pressure is, in general, difficult to obtain from the interior solutions \cite{St}. However, our solution has the nice feature of providing a barotropic EOS. To demonstrate this, we first note from (\ref{37c}) that the variable  $x$ can be expressed completely in terms of the
energy density $\rho$ in the form
\be \label{48}x=\frac{2(C-\rho)\pm \sqrt{2C(11\rho+2C)}}{4\rho}=f(\rho),\ee where $f(\rho)$ denotes the function of $\rho$. \\
Consequently, the pressure $p$ in (\ref{37d}) can be obtained in terms of density $\rho$ as
\be p = C\frac{f_1(\rho)}{f_2(\rho},\n\ee
where
$f_1=a_{1}[2-f(\rho)]^{\frac{1}{2}}[1+2f(\rho)]^{\frac{1}{2}}[19+18f(\rho)-40f(\rho)^{2}]\\
+a_{2}[-34-111f(\rho)-96f(\rho)^{2}+80f(\rho)^{3}],$\\
$f_2=2\{[2-f(\rho)]^{\frac{1}{2}}[1+2f(\rho)]^{2}[a_{1}[1+2f(\rho)]^{\frac{3}{2}}+a_{2}[2-f(\rho)]^{\frac{1}{2}}[7+4f(\rho)]\}.$

\section{Discussion}
In this paper,  we have presented a new technique to generate a large family of solutions to the EM system by making use of Durgapal and Bannerji \cite{DuBa} transformation equations together with a particular fall-off behaviour of the electric field intensity. A large class of solutions obtained previously have been shown to be contained in our general class of solutions. Moreover, we have demonstrated that for the specific set of model parameters, it is possible to obtain closed-form solutions from the general series solution. Note that different families of the solutions depend crucially on the transformation (\ref{eq:12}). It should be stressed here that even though the integral forms in (\ref{eq:32}) and (\ref{eq:33}) parametrized by $d$ do not permit one to regain the previously obtained charge independent solutions, it can be done at a later stage. Once the integrations in (\ref{eq:32}) and (\ref{eq:33}) are obtained in closed forms for specific model parameters, the charged and uncharged solutions in terms of elementary functions can be obtained independently. This technique has been used to regain the Durgapal and Bannerji \cite{DuBa} and Maharaj and Mkhwanazi \cite{MaMk} stellar solutions and also John and Maharaj \cite{JoMa} charged fluid solution. We have also provided two separate closed-form solutions to the EM system. The simple form of the solution facilitates the analysis of the physical behaviour of a charged fluid sphere effectively. Interestingly, our solution has also been shown to provide a barotropic equation of state. It should, however, be pointed out that we have generated closed-form solutions for some specific values of the model parameters. It will be interesting to explore the possibility of generating new class solutions for values of model parameters which have not been covered in this paper. This, however, will be taken up elsewhere.

\section*{Acknowledgement}
KK would like to thank the South Eastern University of Sri Lanka for financial support. RS acknowledges support from the Inter-University Centre for Astronomy and Astrophysics (IUCCA), Pune, India, under its Visiting Research Associateship Programme.




\begin{thebibliography}{99} 
\bibitem{Pap}
Papapertrou A., \textit{Proc. R. Irish. Acad}., \textbf{51} 191,
1947.
\bibitem{Maj}
Majumdar S. D., \textit{Phys. Rev}., \textbf{72} 390, 1947.
\bibitem{Bon1}
Bonner W. B., \textit{J. Phys}., \textbf{59}, 160, 1960.
\bibitem{Bon2}
Bonner W. B., \textit{Mon. Not. R. Astron. Soc}., \textbf{29},
443, 1965.
\bibitem{Ste}
Stettner R., \textit{Ann. Phys}., \textbf{80}, 212, 1973.
\bibitem{Bek}
Bekenstein J. D., \textit{Phys. Rev. D}., \textbf{4}, 2185,
1971.
\bibitem{Coo}
Cooperstock  F. I. and Cruz V. de la,. \textit{Gen. Relativ.
Grav}., \textbf{9}, 835, 1978.
\bibitem{Iv}
 Ivanov B. V.,  \textit{Phys. Rev. D}, \textbf{65} 104001,  2002.
\bibitem{KoMa1}
 Komathiraj K.  and   Maharaj S. D.,  \textit{J. of Math. Phys}.,
\textbf{48},  042501, 2007.
\bibitem{KoMa2}
 Komathiraj K and Maharaj S. D., \textit{Mathematical and Computational
Applications}, \textbf{15},  665, 2010.
\bibitem{PaKo}
 Patel L. K. and     Koppar S. K.,  \textit{Aus.
J. of Phys},  \textbf{40},  441, 1987.
\bibitem{TiSi}
 Tikekar R.  and   Singh G. P.,  \textit{Gravitation and
Cosmology},  \textbf{4}, 294, 1998.
\bibitem{GuKu}
 Gupta Y. K. and   Kumar M.,  \textit{Gen. Relativ.
Grav},   \textbf{37},  233, 2005.
\bibitem{ShMuMh}
Sharma R., Mukherjee S. and Maharaj S. D., \textit{Gen. Relativ. Grav.}, \textbf{33}, 999, 2001.
\bibitem{ShKaMu}
Sharma R., Karmakar S. and  Mukherjee S., \textit{Int. J. Mod.
Phys. D}, \textbf{15}, 405, 2006.
\bibitem{ShMu1}
Sharma R. and  Mukherjee S., \textit{Mod. Phys. Lett. A},
\textbf{16}, 1049, 2001.
\bibitem{ShMu2}
Sharma R. and  Mukherjee S., \textit{Mod. Phys. Lett. A},
\textbf{17}, 2535, 2002.
\bibitem{ThRaVi}
Thomas V.O., Ratanpal B.S. and  Vinodkumar, P.C.,  \textit{Int. J.
Mod. Phys. D},  \textbf{14}, 85, 2005.
\bibitem{TiTh}
Tikekar R. and  Thomas V.O.,  \textit{Prammana - j. of Phys.},
\textbf{50}, 95, 1998.
\bibitem{PaTi}
Paul B.C. and  Tikekar R.,  \textit{Gravitation and Cosmology},
\textbf{11}, 244, 2005.
\bibitem{MaHa}
Mak M.K. and  Harko T., \textit{Int. J. Mod. Phys. D},
\textbf{13}, 149, 2004.
\bibitem{KoMa3}
Komathiraj K. and  Maharaj S.D., \textit{Int. J. Mod. Phys. D},
\textbf{16}, 1803, 2007.
\bibitem{ThMa1}
Thirukkanesh S. and  Maharaj S.D., \textit{Class. Quantum. Grav.}.
\textbf{25}, 35001, 2008.
\bibitem{Var}
Varela V., Rahaman F., Ray S., Chakraborty K. and  Kalam M.,
\textit{Phys. Rev. D}, \textbf{82},
 044052, 2010.
\bibitem{TaMa}
 Takisa P. M. and  Maharaj S.D., \textit{Astrophys. Space. Sci.}, \textbf{343},
569, 2013.
\bibitem{FeSi}
Feroze T. and  Siddique A.A.,  \textit{Gen. Relativ. Grav.},
\textbf{43}, 1025, 2011.
\bibitem{MaTa}
Maharaj S.D. and   Takisa P. M., \textit{Gen. Relativ. Grav.},
\textbf{44}, 1419, 2012.
\bibitem{ThRa1}
Thirukkanesh S. and  Ragel F.C., \textit{Pramana - j. of Phys.},
\textbf{81}, 275, 2013.
\bibitem{ThRa2}
Thirukkanesh S. and  Ragel, F.C., \textit{Pramana - j. of Phys.}
\textbf{78}, 687, 2012.
\bibitem{MaMa}
 Maharaj S.D. and Matondo D. K. and Takisa P. M.,  \textit{Int. J. Mod. Phys. D } \textbf{26}, 1750014,
 2017.
\bibitem{FiSk}
Finch M.R. and  Skea J.E.F., \textit{ Class. Quantum Grav.},
\textbf{6}, 467, 1989.
\bibitem{MuFa}
Murad M.H. and  Fatema S., \textit{Eur. Phys. J. C}, \textbf{75},
533, 2015.
\bibitem{Hans_JMP}
Hansraj S., Maharaj S. D., Mlaba S. and Qwabe N., \textit{J. Math. Phys.}, \textbf{58}, 052501, 2017.
\bibitem{Sunzu}
J. M. Sunzu and P. Danford, \textit{Pramana-j. of physics}, {\bf89}, 44 (2017).
\bibitem{MaLe}
Maharaj S.D. and  Leach P.G.L.,  \textit{J. Math. Phys.}, \textbf{37}, 430,
 1996.
\bibitem{Tik}
Tikekar R., \textit{J. Math. Phys.}, \textbf{31}, 2454, 1990.
\bibitem{DuBa}
 Durgapal M.C. and  Bannerji R., \textit{Phys. Rev. D}, \textbf{27},
 328, 1983.
 \bibitem{MaKo}
 Maharaj S.D. and  Komathiraj K., \textit{Class. Quantum Grav.}, \textbf{24},
4513, 2007.
\bibitem{HaMa}
Hansraj S. and  Maharaj S.D., \textit{Int. J. Mod. Phys. D},
\textbf{15}, 1311, 2006.
\bibitem{TaMa2}
Takisa P. M and  Maharaj S.D., \textit{Gen. Relativ. Grav.}
\textbf{45}, 1951, 2013.
\bibitem{JoMa}
John A.J. and  Maharaj S.D., \textit{Pramana - j. of Phys.},
\textbf{77}, 461, 2011.
\bibitem{MaMk}
Maharaj S.D. and Mkhwanazi W.T., \textit{Questiones Mathematicae.}
\textbf{19}, 211, 1996.
\bibitem{KoMa4}
Komathiraj K. and  Maharaj S.D., \textit{Gen. Relativ. Grav.},
\textbf{39}, 2079, 2007.
\bibitem{DeLa}
Delgaty M. S. R. and Lake K., \textit{Comput. Phys. Commun.},
\textbf{115} 395, 1998.
\bibitem{St}
Stephani H., Kramer D., MacCallum M.A.H., Hoenselaers C. and Herlt
E., \textit{Exact Solutions of Einstein's Field Equations},
Cambridge University Press, Cambridge, 2003.

\end{thebibliography}
\end{document}